\documentclass[aps,prl,superscriptaddress,reprint]{revtex4-1}
\usepackage{graphicx}
\usepackage{amsmath}
\usepackage{amssymb}
\usepackage{amsfonts}
\usepackage{filecontents}
\usepackage{color}
\usepackage[normalem]{ulem} 
\usepackage{soul}
\usepackage{epstopdf}
\usepackage{fancyhdr}
\usepackage{float}
\begin{document}

\title{Dynamic vibronic coupling in InGaAs quantum dots}

\author{A.J.~Brash}
\author{L.M.P.P.~Martins}
\affiliation{Department of Physics and Astronomy, University of Sheffield, Sheffield, S3 7RH, United Kingdom}
\author{A.M.~Barth}
\affiliation{Institut f\"{u}r Theoretische Physik III, Universit\"{a}t Bayreuth, 95440 Bayreuth, Germany}
\author{F.~Liu}
\affiliation{Department of Physics and Astronomy, University of Sheffield, Sheffield, S3 7RH, United Kingdom}
\author{J.H.~Quilter}
\affiliation{Department of Physics and Astronomy, University of Sheffield, Sheffield, S3 7RH, United Kingdom}
\affiliation{Department of Physics, Royal Holloway, University of London, Egham, TW20 0EX, United Kingdom}
\author{M.~Gl\"{a}ssl}
\author{V.M.~Axt}
\affiliation{Institut f\"{u}r Theoretische Physik III, Universit\"{a}t Bayreuth, 95440 Bayreuth, Germany}
\author{A.J.~Ramsay}
\affiliation{Hitachi Cambridge Laboratory, Hitachi Europe Ltd., Cambridge CB3 0HE, United Kingdom}
\author{M.S.~Skolnick}
\affiliation{Department of Physics and Astronomy, University of Sheffield, Sheffield, S3 7RH, United Kingdom}
\author{A.M.~Fox}\email[Corresponding author: ]{mark.fox@sheffield.ac.uk}
\affiliation{Department of Physics and Astronomy, University of Sheffield, Sheffield, S3 7RH, United Kingdom}

\begin{abstract}

The electron-phonon coupling in self-assembled InGaAs quantum dots is relatively weak at low light intensities, which means that the zero-phonon line in emission is strong compared to the phonon sideband. However, the coupling to acoustic phonons can be dynamically enhanced in the presence of an intense optical pulse tuned within the phonon sideband. Recent experiments have shown that this dynamic vibronic coupling can enable population inversion to be achieved when pumping with a blue-shifted laser and for rapid de-excitation of an inverted state with red detuning. In this paper we confirm the incoherent nature of the phonon-assisted pumping process and explore the temperature dependence of the mechanism. We also show that a combination of blue- and red-shifted pulses can create and destroy an exciton within a timescale $ \sim 20$\,ps determined by the pulse duration and ultimately limited by the phonon thermalisation time. 
\end{abstract}

\maketitle

\section{Introduction}

Vibronic sidebands are observed in the optical spectra of many solid-state materials~\cite{fox OPS}, with Ti:sapphire \cite{Klein Nature 2010} and nitrogen-vacancy (NV) centres in diamond \cite{NV ref 1, NV ref  3, NV ref 2} being good examples. The key point about a vibronic transition is that it involves the simultaneous absorption or emission of a photon and a phonon (or phonons)  as the electron jumps between two electronic states. The sideband spectra are continuous bands, although sub-structure can frequently be identified due to the involvement of specific phonon modes, especially at low temperatures. For example, in the case of NV centres in diamond, clear structure can be identified that corresponds to the coupling of the electronic state to the A$_1$ mode ($\hbar \omega =65$~meV), giving rise to resolved side-peaks in both absorption and emission at integer multiples of the phonon energy~\cite{NV ref 1, NV ref  3}. The coupling to the vibrational modes is so strong that only a few percent of the emission occurs in the zero-phonon line, with most of the photons emitted from the sidebands~\cite{NV ref 2}. The relatively weak intensity of the zero-phonon line has serious consequences for practical applications of NV centres in optical quantum information processing.

The reason why the phonon sidebands in materials like diamond NV centres and Ti:sapphire are so strong is that both the electronic and vibrational modes are strongly localized on length scales similar to the unit cell size. This means that the overlap between the electronic wave functions and the phonon modes is large, and hence the vibronic coupling strong. By contrast, InGaAs quantum dots (QDs) have envelope wave functions localized on much larger length scales that are determined by the size of the dot, i.e. $ \sim 10$\,nm. This means that the coupling to phonons is relatively weak, with the dominant interaction being to longitudinal-acoustic (LA) phonons via deformational potential scattering. The weak vibronic coupling in these dots gives rise to very strong emission in the zero-phonon line, with only $ \sim $8\%\ in the sideband at cryogenic temperature~\cite{dot zero-phonon line1, dot zero-phonon line2}. This makes InGaAs QDs excellent single-photon sources~\cite{single photon sources}. It also ensures that the light-matter coupling is strong, with optical dipole moments typically in the range of $\sim 30$~Debye ($\sim 1 \times 10^{-28}$~Cm)~\cite{dot dipole moment}. Moreover, the relatively weak electron-phonon coupling leads to long coherence times that are ultimately limited only by the radiative lifetime~\cite{Langbein} which facilitates their application in coherent control experiments~\cite{dot coherent control, dot coherent control2, dot coherent control3, Ardelt PRB 2014}. This contrasts with bulk and quantum well samples, where the coherence time is only a few picoseconds at best~\cite{fast dephasing in bulk, fast dephasing in bulk book, shah book}.

The vibronic sideband in InGaAs quantum dots manifests itself in a number of important experimental situations. One example is the emission of photons from a nano-cavity when the dot and cavity are out of resonance, with the emission involving either the absorption or emission of a phonon depending on the sign of the detuning~\cite{Off-resonant cavity emission, single photon sources}. Another case is the observation of QD emission when pumping in the phonon sideband in a resonance fluorescence geometry~\cite{Weiler}. The difference in the frequency of the pump laser and the QD exciton line facilitates spectral selection of the QD photons~\cite{Lodahl sideband} and can be exploited for stabilizing the frequency of the zero-phonon line~\cite{Atature sideband}. The sideband also appears in four-wave mixing experiments~\cite{Borri 2005}.

A key point about the experiments described above is that the optical pumping is relatively weak, so that the QD-laser system is in the weak-coupling limit. In this paper we explore the other limit where the QD-laser coupling is strong, such that a dressed-state picture is appropriate. This limit was addressed in a theoretical paper in 2013 by Gl\"{a}ssl \textit{et al}, where it was predicted that strong driving in the phonon sideband could lead to exciton populations approaching full inversion~\cite{Glassl PRL 13}.  The predictions of the theory were confirmed independently by three experimental groups in 2014--15, both for pumping of the neutral exciton and the biexciton~\cite{Quilter PRL 2015, Ardelt PRB 2014, Bounouar PRB 2015}. Most recently, it has also been demonstrated that the process works in reverse, so that an inverted system can be rapidly depopulated by pumping with a laser with red detuning relative to the exciton~\cite{de-excitation}. In these experiments the phonon coupling becomes strong when an intense laser field is present. The pulse durations in the experiments are short, and so the vibronic coupling is turned from weak to strong and back to weak again on picosecond timescales. We therefore call this process \textit{dynamic vibronic coupling} (DVC). The DVC is fundamentally different from conventional excitation schemes, such as coherent Rabi oscillation \cite{Ramsay PRL 2010, Monniello PRL 2013, Ramsay JAP 2011} and adibatic rapid passage protocols \cite{Wei NanoLetters 2014, Mathew PRB 2014} where exciton-phonon coupling is usually an obstacle.

In this paper we first review the process underlying the DVC and summarise the results from our previous experiments~\cite{Quilter PRL 2015, de-excitation}. After discussing the experimental methods in Section~\ref{sec:experiment}, we then present Ramsey interference data to demonstrate that the mechanism of DVC is incoherent, i.e. that the exciton created by phonon-sideband pumping is incoherent with the pumping laser (Section~\ref{sec:coherence}). We next consider the temperature dependence of the DVC in Section~\ref{sec:T-dependence}, comparing the population generated at $\sim 15$\,K to that at base temperature (4.2\,K). Finally we show in Section~\ref{sec:push me pull you} how we can combine pumping with a blue-detuned and a red-detuned pulse to create and destroy an exciton within $\sim 20$\,ps. 
Section~\ref{sec:conclusions} gives the conclusions and outlook.

\section{Mechanism of dynamic vibronic coupling}
\label{sec:mechanism}

\begin{figure}[htbp]
	\centering
	\includegraphics[width=\linewidth]{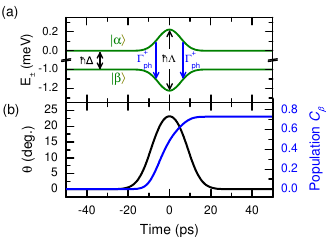}
	\caption{Calculation of the dynamics of the dressed states when excited with a 16.8 ps laser pulse centred at $t$ = 0 with $\hbar\Delta$ = +1~meV and $\Theta$ = 8.5~$\pi$. (a) Energies of the dressed states (green lines) $\left|\alpha\right\rangle$ and $\left|\beta\right\rangle$ plotted against time. At $ t \rightarrow -\infty $, the states are split by the laser detuning $\hbar\Delta$. During the passage of the pulse the states are admixed and the splitting becomes $\hbar\Lambda$ as shown by the shift centred at $t$ = 0. This enables relaxation from $\left|\alpha\right\rangle$ to $\left|\beta\right\rangle$ (blue arrows) by emission of an LA phonon with energy $\hbar\Lambda$. (b) Plot of the admixing angle $\theta$ (black line) and the population of the dressed state $\left|\beta\right\rangle$ (blue line) against time. $\theta$ also follows the envelope of the laser pulse and, as expected, relaxation into $\left|\beta\right\rangle$ occurs only when the states are admixed.}
	\label{fig:mixingJ}
\end{figure}

The starting point for DVC is the coupling of excitons to the acoustic phonon bath by the deformation potential \cite{Krummheurer2002, Barth ArXiv 2016}. In the absence of a laser field this leads to nonexponential pure dephasing of the excitonic dipole ~\cite{Borri 2005}. The behaviour of the coupled system becomes somewhat more interesting however when a strong laser field is applied. We consider a laser pulse with energy detuning $ \hbar \Delta $ and area $\Theta$:
\begin{equation}
\Theta = \int_{-\infty}^{+\infty} \Omega_{\textrm{R}}(t) \, \textrm{d}t \, ,
\end{equation}
where $\Omega_{\textrm{R}} (t)$ is the time-dependent Rabi frequency determined by the optical dipole moment and the time-varying electric field amplitude of the pulse. 
Using the rotating wave approximation, the Hamiltonian in the rotating frame becomes
time independent and one can thus define the dressed states of the laser-QD states
as the corresponding eigenstates. These are given by: 
\begin{equation}
\left|\alpha\right\rangle =\sin\left(\theta\right)\left|0\right\rangle +\cos\left(\theta\right)\left|X\right\rangle
\end{equation}
\begin{equation}
\left|\beta\right\rangle =\cos\left(\theta\right)\left|0\right\rangle -\sin\left(\theta\right)\left|X\right\rangle,
\end{equation}
where $\theta$ is an admixing angle defined by:
\begin{equation}
\label{eq:admixture}
\tan\left(2\theta\right)=-\Omega_{\textrm{R}}/\Delta,\;\;0\leq2\theta\leq180^{\circ}.
\end{equation}
The energies of the dressed states are given by:
\begin{equation}
E_{\pm}=\frac{\hbar}{2}\left(-\Delta \pm \Lambda(t) \right),
\end{equation}
where $\Lambda(t)$ is an effective Rabi frequency defined as:
\begin{equation}
\Lambda(t)=\sqrt{\Omega_{\textrm{R}}(t)^{2}+\Delta^{2}}.
\end{equation}
The significance of these dressed states is that, in the presence of a driving laser such that $\theta>0$, both eigenstates of the system contain an excitonic component and may couple to the acoustic phonon bath. This enables phonon-mediated relaxation from $\left|\alpha\right\rangle$ to $\left|\beta\right\rangle$. However, this relaxation is only possible when the states are admixed by the laser, giving rise to the dynamic nature of the vibronic coupling.

Figure.~\ref{fig:mixingJ} illustrates the dressed state energies (green lines), admixture angle $\theta$ (black line) and occupancy of $\left|\beta\right\rangle$ (blue line) against time during the passage of a $\Theta=8.5\pi$ pulse with $\hbar\Delta=1~\mathrm{meV}$. The pulse has a temporal FWHM of 16.8 ps and is centred on $t=0$. The figure illustrates that the dressed state splitting rises and falls in line with the envelope of the laser pulse~\cite{boyle PRL}, and that the admixing angle follows it. The transfer of population into $\left|\beta\right\rangle$ only occurs whilst the states are admixed, turning the system from weakly to strongly vibronic and back again in approximately 20 ps.

\begin{figure}[htbp]
	\centering
	\includegraphics[width=\linewidth]{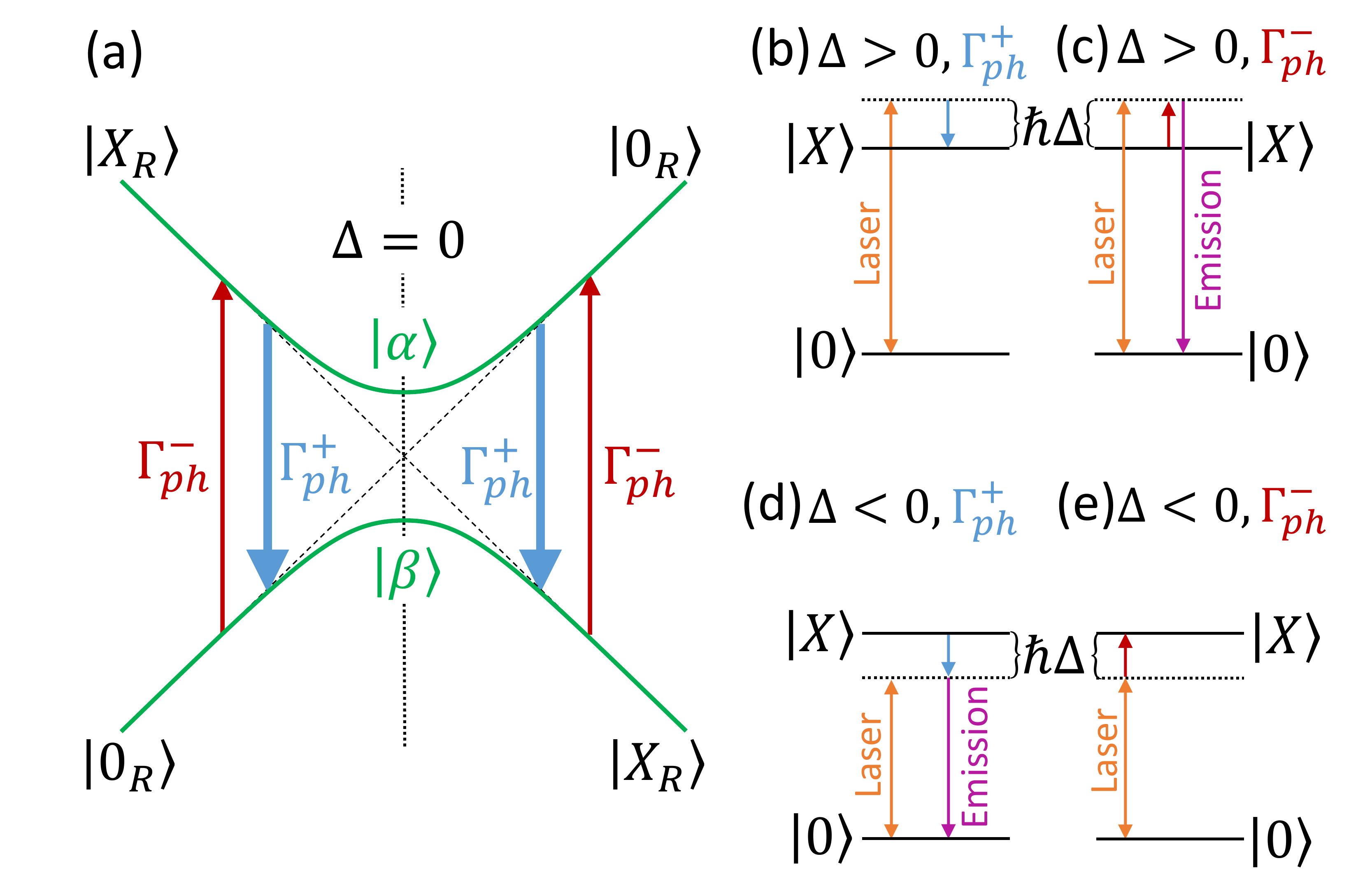}
	\caption{(a) Illustration of the energies of the dressed states $\left|\alpha\right\rangle$ and $\left|\beta\right\rangle$ plotted against detuning $\Delta$ at a fixed time. The bare QD states in the frame of the laser are plotted for reference as dotted black lines. At $\Delta=0$ the states anti-cross with a splitting of $\hbar\Omega_{R}$. It can be seen that  $\left|\alpha\right\rangle$ is dominated by the ground state at $\Delta>0$, leading to phonon-assisted excitation of the exciton. Conversely,  $\left|\alpha\right\rangle$ is primarily excitonic in character at $\Delta<0$, and thus phonon-assisted de-excitation of the exciton occurs. The blue arrows illustrate the relaxation process corresponding to the emission of an LA phonon whilst the red arrows show the competing phonon absorption process. (b-e) Phenomenological energy level diagrams of the phonon emission and absorption processes for both positive and negative detunings. For simplicity, the phonon relaxation is incorporated as a virtual state (dashed lines).}
	\label{fig:relaxation}
\end{figure}

The transfer between the dressed states shown in Fig.~\ref{fig:mixingJ} can be exploited to achieve ultrafast incoherent excitation and de-excitation of the exciton by detuned laser pulses. Beginning with excitation, in the case of $\Delta>0$ the higher energy dressed state $\left|\alpha\right\rangle$ is dominated by the crystal ground state whilst the lower energy state $\left|\beta\right\rangle$ is primarily excitonic in character as illustrated to the right of Fig.~\ref{fig:relaxation}(a). During the passage of the laser pulse, admixing of the states allows relaxation to occur from $\left|\alpha\right\rangle$ to $\left|\beta\right\rangle$ by emission of an LA phonon ($\Gamma_{ph}^{+}$ - blue arrows).

For a sufficiently strong pulse, most of the population relaxes into $\left|\beta\right\rangle$ and the exciton-dominated nature of $\left|\beta\right\rangle$ means that the population of the exciton state after the passage of the pulse is high. Fig.~\ref{fig:relaxation}(b) shows a phenomenological level diagram of the phonon-assisted excitation process whilst Fig.~\ref{fig:relaxation}(c) illustrates the competing process whereby a phonon is absorbed ($\Gamma_{ph}^{-}$ - red arrows) and a photon emitted (purple arrow). The latter process is weak at low temperatures.

In the case of $\Delta<0$ the main difference is that the characteristics of the dressed states are now exchanged. During the passage of the laser pulse, relaxation by phonon emission again occurs into the $\left|\beta\right\rangle$ state. However, as this state is now dominated by the crystal ground state, the effect of the relaxation is instead to de-excite the exciton with most of the population left in the ground state after the passage of the laser pulse. Fig.~\ref{fig:relaxation}(d) shows an illustrative schematic of this process with the emission of a photon (purple arrow) accounting for the rest of the energy difference. This emission is stimulated by the laser and hence the process is termed LA-phonon stimulated emission (LAPSE). Fig.~\ref{fig:relaxation}(e) again illustrates the competing phonon absorption process.

\begin{figure}[htbp]
	\centering
	\includegraphics[width=\linewidth]{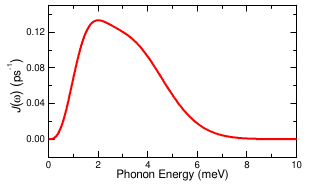}
	\caption{Plot of the exciton-phonon interaction strength $J\left(\omega\right)$ as determined by fitting data measured in Ref.~\cite{Quilter PRL 2015}. The interaction strength peaks at a cut-off of around 2 meV and rolls off rapidly beyond this.}
	\label{fig:phononJ}
\end{figure}

The physical factors that determine the efficiency of phonon-assisted excitation and de-excitation may be considered by analysing the parameters that enter eq.~\ref{eq:admixture}. The degree of admixing increases by increasing the driving strength ($\Omega_{\textrm{R}}$) or decreasing the detuning ($\Delta$) (noting that $\Delta$ should exceed the laser linewidth to exclude resonant coherent driving), leading to more efficient relaxation. However, both of these dependencies are modified by the properties of the phonon bath. The exciton-phonon interaction strength is characterised by the function $J\left(\omega\right)$ which increases with $\omega$ at first due to the rising phonon density of states, and then rolls-off rapidly beyond a cut-off frequency which is typically around 1-2 meV~\cite{dot coherent control2}. The physical origin of this cut-off is the point at which the phonon wavelength is comparable to the spatial FWHM of the carrier wavefunction. As a result, the cut-off frequency depends strongly on the height of the QD and phonon sideband measurements may be used to probe the confinement potential. A plot of $J\left(\omega\right)$ derived from fitting data measured in Ref.~\cite{Quilter PRL 2015} is shown in Fig.~\ref{fig:phononJ}. The form of $J\left(\omega\right)$ modifies the detuning dependence of the DVC and also weakens the exciton-phonon interaction for very strong driving where the effective Rabi frequency $\Lambda$ exceeds the cut-off frequency (although this regime has not been reached in present experiments).

The processes of interest for DVC both rely on phonon emission rather than phonon absorption, and are therefore 
highly sensitive to temperature. 
Increasing the bath temperature increases the phonon occupation leading to a higher probability of phonon emission and thus faster relaxation. However, the probability of phonon absorption generally increases more than that of emission. As illustrated by the red arrows in Fig.~\ref{fig:relaxation}(a), phonon absorption ($\Gamma_{ph}^{-}$) causes the opposite population transfer between the dressed states to phonon emission ($\Gamma_{ph}^{+}$ - blue arrows) and results in a lower final occupation of $\left|\beta\right\rangle$.  As such, DVC processes are generally optimised by low bath temperatures in order to maximise the difference between phonon emission and absorption rates. The influence of temperature on phonon-assisted excitation is studied both experimentally and theoretically in Section~\ref{sec:T-dependence}.

It is only by combining all of these influences that the full spectral and power dependence of the DVC emerges. Analytical approximations do not accurately predict these dependencies; hence numerical methods such as path integral calculations~\cite{Glassl PRL 13, Quilter PRL 2015} or the master equation formalism~\cite{Ardelt PRB 2014} are employed. The result of such calculations is a broad (few meV) spectral sideband feature that appears at high driving strengths and persists until the Rabi splitting exceeds the cut-off energy for phonon coupling.

\section{Experimental methods}
\label{sec:experiment}

\begin{figure}[htbp]
	\centering
	\includegraphics[width=\linewidth]{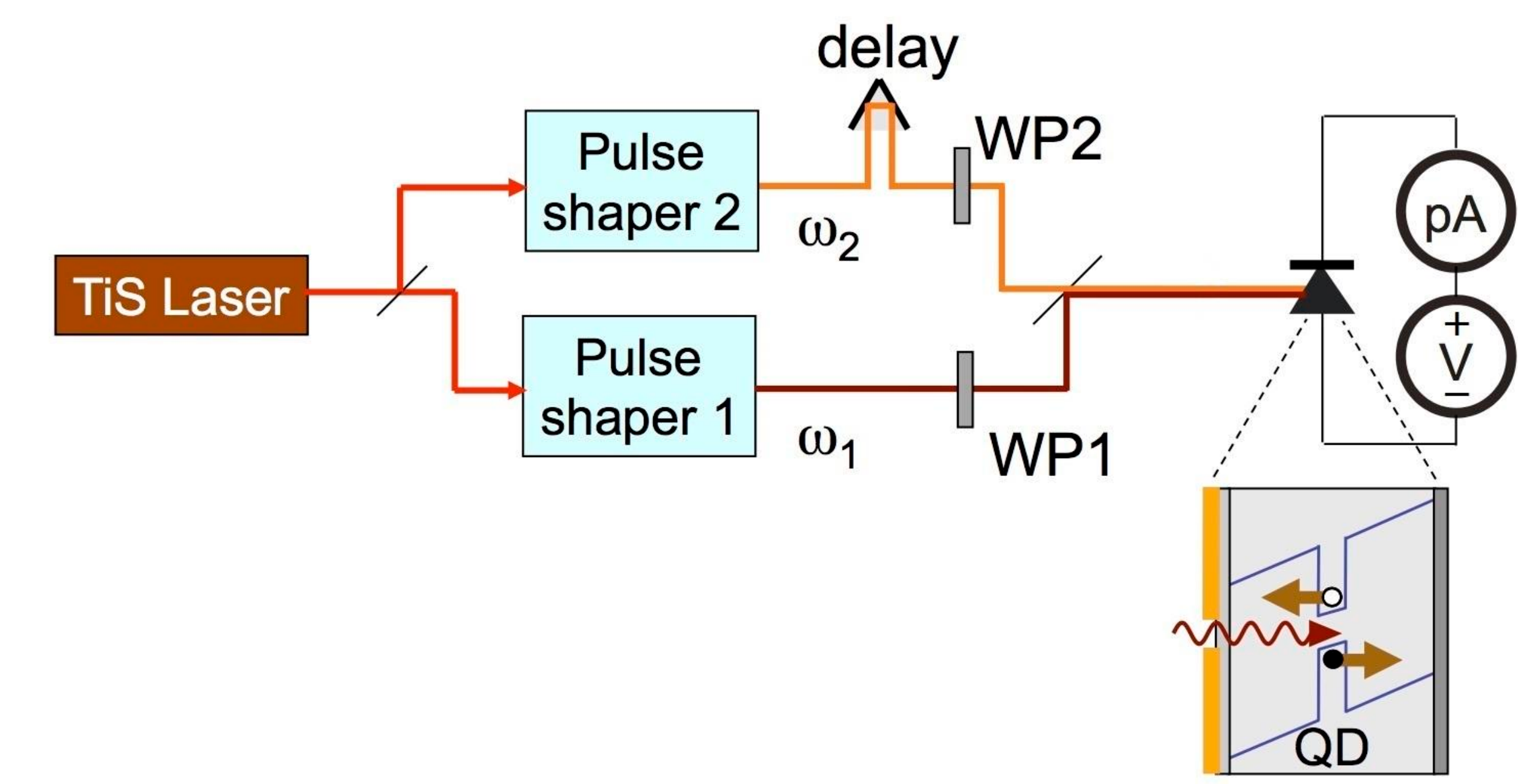}
	\caption{Experimental arrangement. A femtosecond pulsed laser beam is split into two paths. The wavelength, FWHM, arrival time and polarization of the two pulses are controlled independently by the pulse shapers, delay stage and wave plates (WP). Finally the two beams are combined together and sent to the sample which is kept in a variable temperature liquid-He cryostat. The exciton created by the laser pulse is measured by detecting the photocurrent generated from the QD.}
	\label{fig:set-up}
\end{figure}

Figure~\ref{fig:set-up} gives a schematic diagram of the apparatus used in the experiments. A mode-locked Ti:sapphire laser with 76~MHz pulse repetition rate is passed through two pulse shapers to obtain independently tunable pulses from within the $\sim 10$\,meV bandwidth of the $\sim 100$\,fs pulses ~\cite{boyle PRB}. The spectral FWHM is selected to be either either 0.2 or 0.42~meV corresponding to a Fourier-transform-limited pulse duration of 16.8 or 8 ps. 
Both pulses pass through wave-plates to permit independent control of their polarisations, and one of them follows a variable path length controlled by a delay stage to enable pump-probe experiments with precise relative time control.

The pulses are incident on the sample in a variable temperature liquid He cryostat with piezo actuators for precise positioning of the sample relative to the focussed beams. The sample consists of a layer of InGaAs QDs embedded in a Schottky diode. The exciton created by the laser pulses can be measured by detecting the photocurrent (PC) from the QD when a reverse bias is applied to the diode \cite{dot coherent control}. Further details of the experimental methods may be found in ref.~\cite{Brash PRB}.

\section{Coherence and spin persistence of dynamic vibronic coupling}
\label{sec:coherence}

In ref.~\cite{Quilter PRL 2015}, Quilter et al. demonstrated the DVC by creating a QD exciton using a slightly blue-shifted laser pulse via emitting an LA phonon. Since it involves emission of a phonon, in theory the DVC should be incoherent, namely the phase of the exciton is random relative to that of the blue-shifted laser pulse. The exciton coherence time in this case is limited by the phonon relaxation time (few ps \cite{McCutcheon2010}). To demonstrate the incoherent nature of the DVC, we performed Ramsey-like interference measurements \cite{Kamada, Stufler, Kolodka} using an unstabilized interferometer. 

The QD is pumped with a blue-detuned pulse with $\hbar \Delta = 1$~meV at $t=0$. The pulse area ($8.4\pi$) is chosen to give an exciton population of $C_X=0.5$. (Note that this is slightly higher than in ref.~\cite{Quilter PRL 2015} due to different laser detunings.) Once the QD has been excited, the state is then probed by a $\pi/2$ pulse resonant with the exciton at $t=\tau_\text{delay}$. The relative phase between the pump and probe is proportional to $\tau_\text{delay}$. If the system is coherent, the probe pulse drives the QD to either $C_X=0$ or 1 depending on the phase of the exciton with respect to the probe. The degree of coherence can then be determined by measuring the visibility of the Ramsey fringes as a function of $\tau_\text{delay}$. In our experiment, the relative phase between the pump and probe is unstabilized, and so the final state fluctuates randomly within the visibility envelope. 

\begin{figure}[!h]
	\centering
	\includegraphics[width=\linewidth]{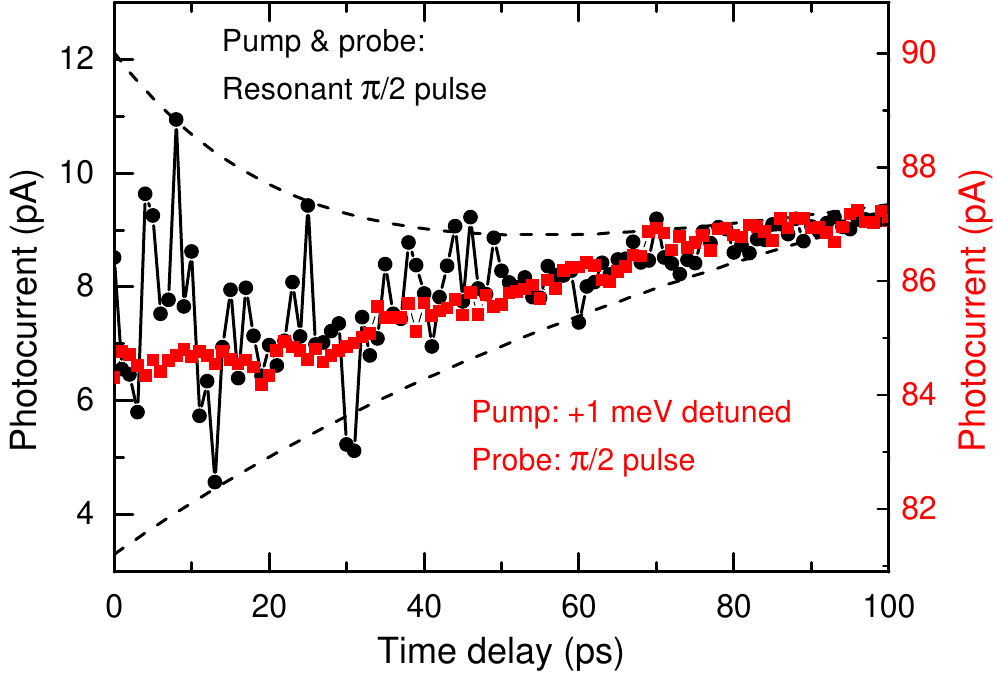}
	\caption{Comparison of Ramsey-like fringe data on the exciton for pumping in the phonon sideband (red) and at the exciton (black) at $T=4.2$\,K. In both cases the second $\pi/2$-pulse is resonant with the exciton. The pulse area $\Theta$ of the +1 meV detuned pump pulse is set to be $8.4\pi$ to generate $C_X = 0.5$. The dashed line is a guide to the eye.}
	\label{fig:Ramsey fringes}
\end{figure}

Figure.~\ref{fig:Ramsey fringes} shows the time-integrated PC signal versus $\tau_\text{delay}$.
The red squares show the results for phonon-sideband pumping with $\hbar \Delta =1$~meV. A flat line is observed, with no Ramsey-like fringes, indicating that the phase of the exciton is random relative to the probe pulse. The slow increase of the PC signal in time is most probably related to the decay of the exciton population created by the pump via electron/hole tunnelling before the arrival of the probe. By contrast, the black dots show the results measured when the pump pulse is tuned to resonance ($\Delta = 0$) and the pulse area is set to $\pi/2$. In this fully resonant situation, Ramsey-like fringes are observed provided $\tau_\text{delay}$ is shorter than or comparable to the exciton coherence time. The exciton coherence time of $\sim 40$~ps can be estimated from the dashed line envelope and is limited by the electron tunnelling rate \cite{Kolodka}. The incoherent nature of the DVC is clearly demonstrated by the absence of the fluctuations in the PC signal in the red data compared with that in the black data. Our result is consistent with that reported by Weiler et al. \cite{Weiler} showing that the coherence properties of the emitted photon created via DVC is much worse than resonant CW excitation. However, Bounouar et al. \cite{Bounouar PRB 2015} demonstrate that the photons generated by quasi-resonantly pumping the biexciton state using pulsed excitation show similar coherence properties as measured in the resonant two-photon scheme. The relation of the coherence properties of the emitted photon and exciton prepared via DVC is still unclear and require further investigations.

It should be noticed that the fact that the DVC process is incoherent does not imply that the exciton spin is random. The results published in ref.~\cite{Quilter PRL 2015} clearly shows that the spin of the exciton is the same as that of the photon in the blue-shifted laser pulse. In theory the spin of the emitted photon in the LAPSE process should also be the same as that of the laser photon, since the phonon-assisted excitation and deexcitation are two opposite processes in the dressed state picture [see Fig.~\ref{fig:relaxation}(a)]. 

\begin{figure}[!h]
	\centering
	\includegraphics[width=\linewidth]{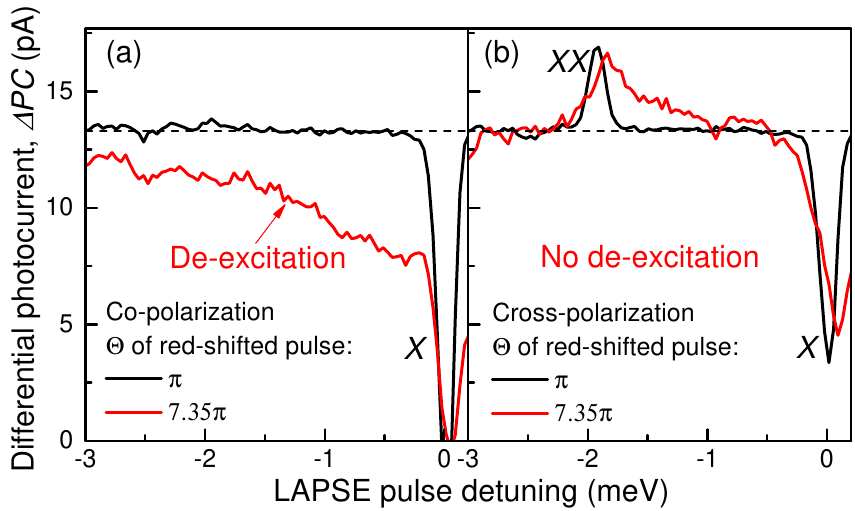}
	\caption{Demonstration of the spin selectivity of the LAPSE process. Differential photocurrent spectra $ \Delta PC$ were obtained by preparing an exciton in the QD using a circularly-polarized $ \pi $ pulse and then measuring the PC as a function of a (a) co- or (b) cross-polarized red-shifted laser pulse. To isolate the PC signal of the QD under study from other QDs in the same sample, a reference spectrum measured with only a red-shifted laser pulse is subtracted. The pulse area $ \Theta $ of the red-shifted laser pulse is (black) $ \pi $ or (red) $ 7.35\pi $. The delay time between the $ \pi $ pulse and the red-shifted pulse is 17~ps. $ XX $: biexciton.}
	\label{fig:spin}
\end{figure}

Directly demonstrating the spin-preserving nature of the LAPSE process is beyond the scope of this paper, but it has been observed that the LAPSE process only occurs when the spin of the exciton and the LAPSE pulse are the same. This spin selectivity of the LAPSE process can be demonstrated by preparing a spin-up exciton using a $ \sigma+ $ circularly-polarized $ \pi $ pulse and then try to deexcite this exciton using a (a) co- or (b) cross-polarized red-shifted laser pulse [see Figure.~\ref{fig:spin}]. Figure.~\ref{fig:spin} shows the differential PC spectra $ \Delta PC $ measured by exciting the QD using a $ \sigma+ $ polarized $ \pi $ pulse and then measuring the exciton population by detecting the PC signal as a function of the detuning of a subsequent red-shifted co/cross-polarized laser pulse. The first $ \pi $ pulse creates a reference PC level corresponding to an exciton population of 1 [see the dashed line in Fig.~\ref{fig:spin}(a)]. This reference level is calibrated by the PC signal at $ \hbar\Delta<0 $ in the differential spectrum measured with a weak ($ \Theta=\pi $) co-polarized red-shifted pulse where the exciton-phonon coupling is negligible since the laser driving is quite weak [see black line in Fig.~\ref{fig:spin}(a)]. In the case that a strong ($ \Theta=7.35\pi $) co-polarized red-shifted pulse is applied, the red-shifted pulse deexcites the exciton, hence reduces the PC single relative to the reference level and forms a negative sideband at $ \hbar\Delta<0 $ [see red line in Fig.~\ref{fig:spin}(a)]. By contrast, no decrease of the PC single is observed at $ \hbar\Delta<0 $ in the cross-polarized case [see red line in Fig.~\ref{fig:spin}(b)], indicating that no de-excitation occurs. The peak and weak positive sideband at around $ \hbar\Delta=-1.96 $~meV can be attributed to the resonant and phonon-assisted excitation of the biexction. The absence of the negative phonon sideband in the cross-polarized differential spectrum compared to the co-polarized spectrum [see red lines in Fig.~\ref{fig:spin}] ambiguously proves the spin selectivity of the LAPSE process. Based on this observation and the fact that the spin is preserved while the emission of the LA phonon as demonstrated in ref. \cite{Quilter PRL 2015}, we conclude that DVC is incoherent, but spin-preserving. 

\section{Temperature dependence of dynamic vibronic coupling}
\label{sec:T-dependence}

\begin{figure}[htbp]
	\centering
	\includegraphics[width=\linewidth]{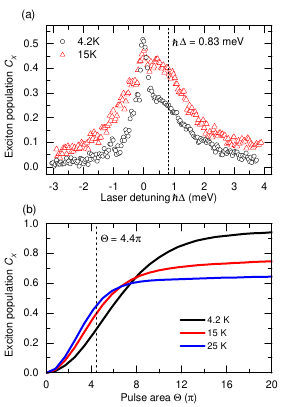}
	\caption{(a) Exciton population vs. laser detuning ($\Delta$) at 15\,K (red triangles) compared to 4.2\,K (black circles) for excitation with pulse area $\Theta$ = 4.4 $\pi$. The dashed line indicates a detuning of $\hbar\Delta$ = +0.83 meV. (b) Calculated variation of the exciton population versus pulse area ($\Theta$) for $T=4.2$\,K (black), $T=15$\,K (red) and $T=25$\,K (blue) for positive detuning of $\hbar\Delta$ = +0.83 meV. The dashed line shows $\Theta$ = 4.4 $\pi$.}
	\label{fig:detuning}
\end{figure}

In previous studies DVC processes have mainly been studied at $T=4.2$\,K \cite{Quilter PRL 2015, Ardelt PRB 2014, Bounouar PRB 2015}, the temperature of liquid-He. It is interesting however to consider the influence of temperature on the mechanism. Increasing the temperature increases the strength of the exciton-phonon interaction, observed for instance by stronger damping and frequency renormalization of excitonic Rabi rotations~\cite{dot coherent control2}. As discussed in section~\ref{sec:mechanism}, this increases the rate of phonon relaxation between the dressed states, and, for relatively weak excitation, increases the resulting exciton population as more population relaxes during the pump pulse. This is shown by the higher population observed at positive detuning for $T=15$\,K (red triangles) compared to $T=4.2$\,K (black circles) in Fig.~\ref{fig:detuning}(a) and also by the steeper initial gradient for higher temperatures observed in the calculations of Fig.~\ref{fig:detuning}(b).

However, this does not illustrate the complete picture. At higher bath temperatures the increased number of phonons increases the probability of the competing process at positive detuning, namely the annihilation of an exciton by absorption of a phonon (see Fig.~\ref{fig:relaxation}(c)). This is shown in our measurements by the reduced detuning asymmetry of the phonon sidebands at $T=15$\,K in Fig.~\ref{fig:detuning}(a). The origin of the negative detuning signal is the creation of exciton population by phonon absorption, illustrating the increased probability of the competing absorption processes discussed in section ~\ref{sec:mechanism}. As such, the reduced asymmetry in the phonon sidebands can be considered equivalent to the reduced difference between the probability of phonon emission and absorption.

The consequence of this competition is that for driving that is sufficiently strong to allow complete relaxation (i.e. in the plateau region of Fig.~\ref{fig:detuning}(b)), the final occupancy of the exciton state is in fact lower at higher temperatures owing to the thermalisation of the dressed states. This is illustrated by the reduced maximum exciton population attained at higher bath temperatures in Fig.~\ref{fig:detuning}(b). In summary, a higher bath temperature will lead to a faster rise with pulse area, but a lower maximum population transfer by DVC. As such, increasing the bath temperature may provide a means to enhance DVC processes when the driving is relatively weak as in Fig.~\ref{fig:detuning}(a).

\section{Picosecond timescale excitation and de-excitation}
\label{sec:push me pull you}

The three first experimental observations of DVC in quantum dots~\cite{Quilter PRL 2015, Ardelt PRB 2014, Bounouar PRB 2015} focused on the LA-phonon-assisted excitation process [see Fig.~\ref{fig:relaxation}(a), (b)]. Recently Liu. et al. have demonstrated the reverse process - LA-phonon-assisted de-excitation [see Fig.~\ref{fig:relaxation}] - by observing the reduction of the exciton population of an excited QD by a red-shifted laser pulse \cite{de-excitation}. A combination of the LA-phonon-assisted excitation and de-excitation processes allows us to create and destroy an exciton within picosecond time scales, enabling ultrafast incoherent optical switching with a single QD much faster than the exciton radiative lifetime ($ \sim $ns \cite{Quilter PRL 2015, Dalgarno}). 

\begin{figure}[htbp]
	\centering
	\includegraphics[width=\linewidth]{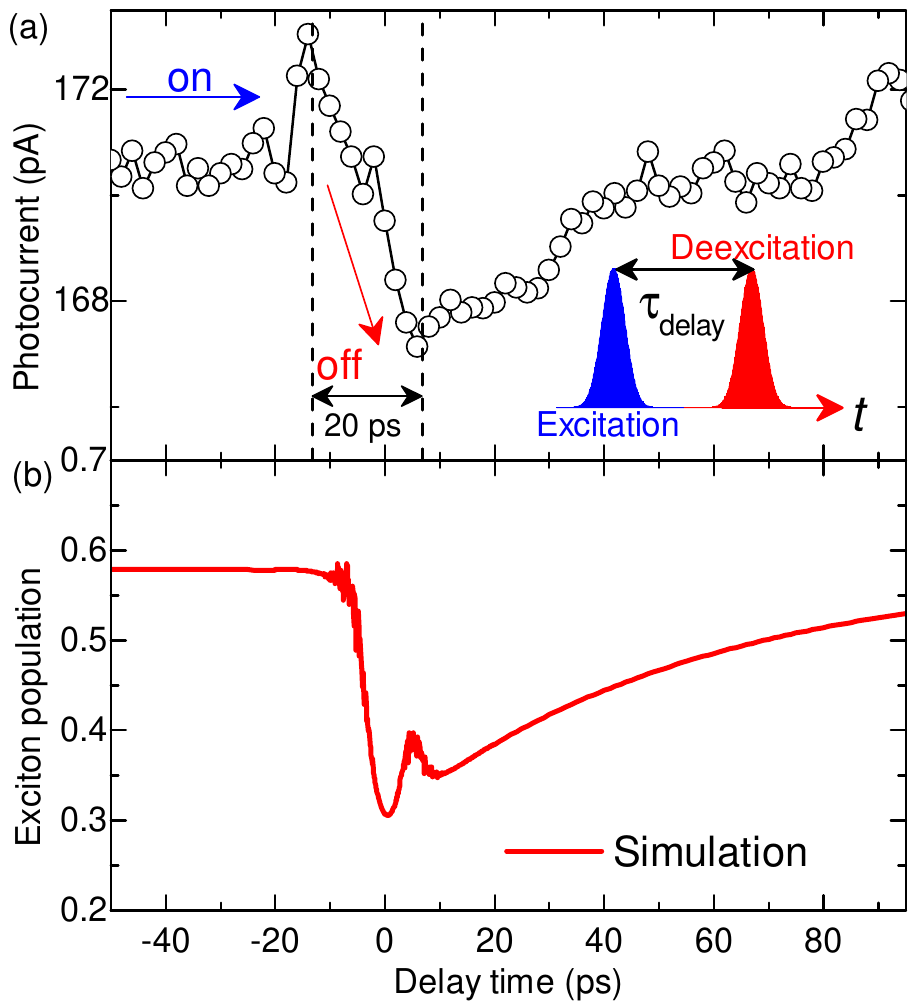}
	\caption{(a) PC signal measured with positively and negatively de-tuned control pulses as a function of the delay time $ \tau_\text{delay} = t_\text{OFF}-t_\text{ON}$, where $ t_\text{OFF}$ and $t_\text{ON} $ are the arrival time of the de-excitation and excitation pulses, respectively. $ \hbar \Delta $ = $\pm$0.7 meV. $\Theta = 5.25\pi $. $ \text{FWHM} = 0.42  $~meV. $ \text{Pulse duration} = 8$~ps. Inset: Pulse sequence. (b) Exciton population detected in the two-color pump-probe measurement calculated using similar parameters as Fig.~\ref{fig:Push-pull}(a). $ \hbar \Delta $ = $\pm$0.7 meV. $\Theta = 5.25\pi $. $ \text{Pulse duration} = 10$~ps.}
	\label{fig:Push-pull}
\end{figure}

To demonstrate this, a two-color pump-probe experiment is performed with a blue-shifted (excitation) and red-shifted (de-excitation) pulse with the same pulse area. The pulses are equally detuned ($ \hbar \Delta=\pm0.7 $~meV) from the exciton. At negative delay times, the excitation pulse creates a certain exciton population corresponding to the background PC level. The de-excitation pulse arrives before excitons are created, and therefore does nothing. When the two pulses overlap, the PC signal decreases to a minimum within $20$ ps, as shown in Fig.~\ref{fig:Push-pull}. This occurs due to erasure of the exciton created by the excitation pulse by the de-excitation pulse. The PC signal then slowly recovers due to the electron tunnelling out from the QD before the de-excitation pulse arrives \cite{de-excitation}. The increase of the PC signal at $ \sim -10 $~ps is most probably related to the phonon-assisted excitation of biexciton by a very small portion of imperfectly circularly polarized light in the excitation/deexcitation pulses. Fig.~\ref{fig:Push-pull}(b) shows the calculated exciton population with similar parameters as Fig.~\ref{fig:Push-pull}(a). The phase relation between the two pulses has been assumed to be random and we show the average of multiple repetitions of the simulation. The numerically complete path-integral method takes into account the electron tunnelling in the PC experiment by incorporating it as an Lindblad-type relaxation term \cite{de-excitation}. Before the arrival of the pulses the QD is assumed to be in the ground-state and the phonon modes follow a thermal distribution. This model well reproduces the deexcitation of the QD and the subsequent recovery of the photo current signal. The oscillation of the exciton population at $ \sim $5 ps is not clearly visible in the measured data in Fig.~\ref{fig:Push-pull}(a) due to the limited time resolution of the experiment.

The $\sim 20$~ps excitation/de-excitation time in our experiment is determined by the laser pulse duration. The ultimate limit, , is set by the phonon thermalisation time (few ps \cite{McCutcheon2010}). We note that in contrast to the optical switching scheme employing resonant coherent excitation~\cite{Heberle}, our scheme is robust against the fluctuation of laser power and detuning~\cite{Glassl PRL 13}. Furthermore, the incoherent nature of the phonon-assisted relaxation process determines that no phase locking  is needed, in contrast to fully coherent excitation schemes.

\section{Conclusions and outlook}
\label{sec:conclusions}

In summary, we have investigated the dynamic vibronic coupling in a single InGaAs QD. The DVC enables the population inversion and ultrfast depopulation of a QD by incoherent quasi-resonant excitation with the assistant of phonons \cite{Quilter PRL 2015, de-excitation}.
%The underlying mechanism can be understood as that a strong laser pulse turns the QD from a weak vibronic system to a strong one by enabling the relaxation between two optically via emitting phonons.
We prove that the DVC is an incoherent process by performing Ramsey-like interference measurements. The temperature dependence study shows that at high temperature the phonon-assisted absorption induced by both the blue- and red-shifted laser pulses are enhanced and the strong asymmetry of the phonon sidebands observed at low temperatures disappears. Furthermore we demonstrate that it is possible to create and destroy an exciton within a pulse-width limited ps time scale by combining a blue-detuned and a red-detuned pulses, opening the door to ultrafast incoherent optical switching with a single QD \cite{Heberle,Volz, Cancellieri}. The DVC may also be used to reduce the timing jitter of single/entangled photon sources \cite{Heinze 2015, single photon sources} or make tunable single QD lasers \cite{Nomura 2010}. 

Compared with resonant coherent excitation, one factor which limits the potential applications of DVC is the high laser power required to achieve an efficient phonon-assisted excitation/deexcitation. This problem may be solved by incorporating the QD into a nanocavity with a small mode volume where the light matter interaction is strongly enhanced \cite{single photon sources, hughes 2013}. When cavities are involved, the knowledge of phonon interactions in the strong QD-phonon coupling regime becomes an essential requirement for a full understanding of the behaviour of optically driven QDs. Various recent studies have shown new physical phenomena that arise from this coupling \cite{Dara 2013, Wei 2014, Muller 2015, Jake 2016, Hargart 2016}.

\section*{Acknowledgements}

This work was funded by the EPSRC (UK) EP/J007544/1. 
M. G., A. M. B., and V. M. A. gratefully acknowledge the financial support from Deutsche Forschungsgemeinschaft via the Project No. AX 17/7-1. The authors thank H.Y. Liu and M. Hopkinson for sample growth.

%Manual citation list


\begin{thebibliography}{1}
	
	\bibitem{fox OPS}
	See, for example, M. Fox, \textit{Optical properties of solids}, 2nd edn (Oxford University Press, 2010).
	
	\bibitem{Klein Nature 2010}
	J. Klein and J. D. Kafka, ``The Ti:Sapphire laser: The flexible research tool'', Nature Photonics 4, 289 (2010).
	
	\bibitem{NV ref 1}
	G. Davies, ``The Jahn-Teller effect and vibronic coupling at deep levels in diamond'', Rep. Prog. Phys., \textbf{44}, 787 (1981).
	
	\bibitem{NV ref  3}
	G. Davies, M. F. Hamer, ``Optical Studies of the 1.945 eV Vibronic Band in Diamond'', Proc. R. Soc. Lond. A., \textbf{348}, 285-298 (1976).
	
	\bibitem{NV ref 2}
	Romana Schirhagl, Kevin Chang, Michael Loretz, and Christian L. Degen, ``Nitrogen-Vacancy Centers in Diamond: Nanoscale Sensors for Physics and Biology'', Annual Review of Physical Chemistry \textbf{65}, 83-105 (2014).
	
	
	
	\bibitem{dot zero-phonon line1} 
	I. Favero, G. Cassabois, R. Ferreira, D. Darson, C. Voisin, J. Tignon, C. Delalande, G. Bastard, and Ph. Roussignol, ``Acoustic phonon sidebands in the emission line of single InAsÕGaAs quantum dots'', Physical Review B \textbf{68}, 233301 (2003)
	
	\bibitem{dot zero-phonon line2} 
	S.L. Portalupi, G. Hornecker, V. Giesz,T. Grange, A. Lema\^{i}tre, J. Demory, I. Sagnes, N.D. Lanzillotti-Kimura, L. Lanco, ,A. Auff\`{e}ves, and P. Senellart, ``Bright Phonon-Tuned Single-Photon Source'', Nano Letters, \textbf{15}, 6290–62945 (2015).
	
	\bibitem{single photon sources} 
	P. Lodahl, S. Mahmoodian, and S. Stobbe, ``Interfacing single photons and single quantum dots with photonic nanostructures'', Rev.  Mod. Phys. \textbf{87}, 347 (2015).
	
	\bibitem{dot dipole moment}
	P. G. Eliseev, H. Li, A. Stintz, G. T. Liu, T. C. Newell, K. J. Malloy and L. F. Lester, ``Transition dipole moment of InAs/InGaAs quantum dots from experiments on ultralow-threshold laser diodes'', Applied Physics Letters \textbf{77}, 262 (2000).
	
	\bibitem{Langbein}
	P. Borri, W. Langbein, S. Schneider, U. Woggon, R.L. Sellin, D. Ouyang, and D. Bimberg,
	``Ultralong Dephasing Time in InGaAs Quantum Dots'',
	Phys. Rev. Lett. \textbf{87}, 157401 (2001).
	
	\bibitem{dot coherent control}
	A. Zrenner, E. Beham, S. Stuer, F. Findeis, M. Bichler, and G. Abstreiter,
	``Coherent properties of a two-level system based on a quantum-dot photodiode'',
	Nature \textbf{418}, 612 (2002).
	
	\bibitem{dot coherent control2}
	A. J. Ramsay, T.M. Godden, S. J. Boyle, E.M. Gauger, A. Nazir, B.W. Lovett, A.M. Fox, and M.S. Skolnick, ``Phonon-Induced Rabi-Frequency Renormalization of Optically Driven Single InGaAs/GaAs Quantum Dots'',
	Phys. Rev. Lett. \textbf{105}, 177402 (2010)
	
	\bibitem{dot coherent control3}
	T. M. Godden, J. H. Quilter, A. J. Ramsay, Y. Wu, P. Brereton, S. J. Boyle, I. J. Luxmoore, J. Puebla-Nunez, A. M. Fox and M. S. Skolnick, ``Coherent Optical Control of the Spin of a Single Hole in an InAs/GaAs Quantum Dot'',
	Phys. Rev. Lett. \textbf{108}, 017402 (2012)
	
	\bibitem{Ardelt PRB 2014}
	P.-L. Ardelt, L. Hanschke, K. A. Fischer, K. M\"{u}ller, A. Kleinkauf, M. Koller, A. Bechtold, T. Simmet, J. Wierzbowski, H. Riedl, G. Abstreiter, and J. J. Finley, ``Dissipative preparation of the exciton and biexciton in self-assembled quantum dots on picosecond time scales'', Phys. Rev. B 90, 241404(R) (2014).
	
	
	\bibitem{fast dephasing in bulk}
	A. P. Heberle and J.J. Baumberg, ``Ultrafast Coherent Control and Destruction of Excitons in Quantum Wells'', Physical Review Letters, \textbf{75}, 2598 (1995)
	
	\bibitem{fast dephasing in bulk book}
	O. Svelto, S. De Silvestri, G. Denardo, \textit{Ultrafast Processes in Spectroscopy} (Springer Science+Business Media, LLC, New York, 1996)
	
	\bibitem{shah book}
	J. Shah, \textit{Ultrafast Spectroscopy of Semiconductors and Semiconductor Nanostructures}, 2nd edition (Springer--Verlag, 1999).
	
	\bibitem{Off-resonant cavity emission}
	S. Ates, S. M. Ulrich, A. Ulhaq, S. Reitzenstein, A. L\"offler, S. H\"ofling,
	A. Forchel and P. Michler, ``Non-resonant dot-cavity coupling and its
	potential for resonant single-quantum-dot-spectroscopy'', Nature Photonics
	\textbf{3}, 724-728 (2009)
	
	\bibitem{Weiler}
	S. Weiler, A. Ulhaq, S. M. Ulrich, D. Richter, M. Jetter, P. Michler, C. Roy and S. Hughes, 
	``Phonon-assisted incoherent excitation of a quantum dot and its emission properties'', Phys. Rev. B \textbf{86}, 241304(R) (2012).
	
	\bibitem{Lodahl sideband}
	K. H. Madsen, S. Ates, J. Liu, A. Javadi, S. M. Albrecht, I. Yeo, S. Stobbe, and P. Lodahl, ``Efficient out-coupling of high-purity single photons from a coherent quantum dot in a photonic-crystal cavity'', Phys. Rev. B \textbf{90}, 155303 (2014)
	
	\bibitem{Atature sideband}
	J. Hansom, C. H. H. Schulte, C. Matthiesen, M. J. Stanley and M. Atat\"{u}re, 
	``Frequency stabilization of the zero-phonon line of a quantum dot via phonon-assisted active feedback'',
	Appl. Phys. Lett. \textbf{105}, 172107 (2014).
	
	\bibitem{Borri 2005}
	P. Borri, W. Langbein, U. Woggon, V. Stavarache,
	D. Reuter, and A. Wieck, ``Exciton dephasing via phonon interactions in InAs quantum dots: Dependence on quantum confinement'', Phys. Rev. B \textbf{71}, 115328
	(2005).
	
	\bibitem{Glassl PRL 13}
	M. Gl\"{a}ssl, A. M. Barth, and V. M. Axt, 
	``Proposed robust and high-fidelity preparation of excitons and biexcitons in semiconductor quantum dots making active use of phonons'',
	Phys. Rev. Lett. \textbf{110},147401 (2013).
	
	\bibitem{Quilter PRL 2015}
	J.H. Quilter, A.J. Brash, F. Liu, M. Gl\"{a}ssl, A.M. Barth, V.M. Axt, A.J. Ramsay, M.S. Skolnick and A.M. Fox, ``Phonon-assisted population inversion of a single InGaAs/GaAs quantum dot by pulsed laser excitation'', Phys. Rev. Lett. \textbf{114}, 137401 (2015).
	
	\bibitem{Bounouar PRB 2015}
	S. Bounouar, M. M\"{u}ller, A. M. Barth, M. Gl\"{a}ssl, V. M. Axt and P. Michler, 
	``Phonon-assisted robust and deterministic two-photon biexciton preparation in a quantum dot'',
	Phys. Rev. B \textbf{91}, 161302 (2015).
	
	\bibitem{de-excitation}
	F. Liu, L.M.P. Martins, A.J. Brash, A.M. Barth, J.H. Quilter, V. M. Axt, M.S. Skolnick, and A.M. Fox
	``Ultrafast depopulation of a quantum dot by LA-phonon-assisted stimulated emission'', Phys. Rev. B \textbf{93}, 161407(R) (2016).
	
	\bibitem{Ramsay PRL 2010}
	A. J. Ramsay, A. V. Gopal, E. M. Gauger, A. Nazir, B.W. Lovett, A. M. Fox, and M. S. Skolnick, ``Damping of Exciton Rabi Rotations by Acoustic Phonons in Optically Excited InGaAs/GaAs Quantum dots'', Phys. Rev. Lett. \textbf{104}, 017402 (2010).
	
	\bibitem{Monniello PRL 2013}
	L. Monniello, C. Tonin, R. Hostein, A. Lemaitre, A.
	Martinez, V. Voliotis, and R. Grousson, ``Excitation-Induced Dephasing in a Resonantly Driven InAs/GaAs Quantum Dot'', Phys. Rev. Lett. \textbf{111}, 026403 (2013).
	
	\bibitem{Ramsay JAP 2011}
	Ramsay, A. J., Godden, T. M., Boyle, S. J., Gauger, A. Nazir, B.W. Lovett, A.V. Gopal,  A.M. Fox, and M. S. Skolnick, ``Effect of detuning on the phonon induced dephasing of optically driven InGaAs/GaAs quantum dots'', Journal of Applied Physics, \textbf{109}, 102415 (2011)
	
	\bibitem{Wei NanoLetters 2014}
	Y.J. Wei, Y.M. He, M.C. Chen, Y.N. Hu, Y. He, D. Wu, C. Schneider, M. Kamp, S. H\"ofling, C.Y. Lu and J.-W. Pan,``Deterministic and Robust Generation of Single Photons from a Single Quantum Dot with 99.5$ \% $ Indistinguishably Using Adiabatic Rapid Passage'', Nano Lett. \textbf{14}, 6515 (2014).
	
	\bibitem{Mathew PRB 2014}
	R. Mathew, E. Dilcher, A. Gamouras, A. Ramachandran, H. Y. Shi Yang, S. Freisem, D. Deppe, and K. C. Hall, ``Subpicosecond adiabatic rapid passage on a single semiconductor quantum dot: Phonon-mediated dephasing in the strong-driving regime'', Phys. Rev. B \textbf{90}, 035316 (2014).
	
	\bibitem{Krummheurer2002}
	B. Krummheuer, V. M. Axt, and T. Kuhn, 
	``Theory of pure dephasing and the resulting absorption line shape in semiconductor quantum dots'',
	Phys. Rev. B \textbf{65}, 195313  (2002).
	
	\bibitem{Barth ArXiv 2016}
	A. M. Barth, S. L\"uker, A. Vagov, D. E. Reiter, T. Kuhn, and V. M. Axt, 
	``Fast and selective phonon-assisted state preparation of a quantum dot by adiabatic undressing'', arXiv:1601.07886 (2016)
	
	\bibitem{boyle PRL}
	S. J. Boyle, A. J. Ramsay, A. M. Fox, M. S. Skolnick, A. P. Heberle and M. Hopkinson, 
	``Beating of exciton dressed states in a single semiconductor InGaAs/GaAs quantum dot'',
	Phys. Rev. Lett. \textbf{102}, 207401 (2009).
	
	\bibitem{boyle PRB}
	S. J. Boyle, A. J. Ramsay, F. Bello, H. Y. Liu, M. Hopkinson, A. M. Fox and M. S. Skolnick,
	``Two-qubit conditional quantum-logic operation in a single self-assembled quantum dot'',  
	Phys. Rev. B \textbf{78}, 075301 (2008).
	
	\bibitem{Brash PRB}
	A. J. Brash, L. M. P. P. Martins, F. Liu, J. H. Quilter, A. J. Ramsay, M. S. Skolnick, and A. M. Fox, ``High-fidelity initialization of long-lived quantum dot hole spin qubits by reduced fine-structure splitting'', Physical Review B \textbf{92}, 121301 (2015).
	
	\bibitem{McCutcheon2010}
	McCutcheon, Dara P S and Nazir, Ahsan, 
	``Quantum dot Rabi rotations beyond the weak exciton–phonon coupling regime'',
	New Journal of Physics, \textbf{12}, 113042 (2010)
	
	\bibitem{Kamada}
	H. Kamada, H. Gotoh, J. Temmyo, T. Takagahara, and H. Ando, 
	``Exciton Rabi Oscillation in a Single Quantum Dot'',
	Phys. Rev. Lett., \textbf{87}, 246401 (2001)
	
	\bibitem{Stufler}
	S. Stufler, P. Ester, A. Zrenner, and M. Bichler, 
	``Exciton Rabi Oscillation in a Single Quantum Dot'',
	Phys. Rev. B, \textbf{72}, 121301(R) (2005)
	
	\bibitem{Kolodka}
	Kolodka, R. S., Ramsay, A. J., Skiba-Szymanska, J., Fry, P. W., Liu, H. Y., Fox, A. M., and Skolnick, M. S., 
	``Inversion recovery of single quantum-dot exciton based qubit'',
	Phys. Rev. B, \textbf{75}, 193306(R) (2007)
	
	\bibitem{Dalgarno}
	P.A. Dalgarno, J.M. Smith, J. McFarlane, B.D. Gerardot, K. Karrai, A. Badolato,
	P.M. Petroff, and R.J. Warburton, 
	``Coulomb interactions in single charged self-assembled quantum dots:
	Radiative lifetime and recombination energy'',
	Phys. Rev. B, \textbf{77}, 245311 (2008)
	
	\bibitem{Heberle}
	A. Heberle, J. Baumberg, and K. K\"ohler, 
	``Ultrafast Coherent Control and Destruction of Excitons in Quantum Wells'',
	Phys. Rev. Lett., \textbf{7P}, 2598 (1995)
	
	\bibitem{Volz}
	T. Volz, A. Reinhard, M. Winger, A. Badolato, K. J. Hennessy, E. L. Hu, and A. Imamoglu, 
	``Ultrafast all-optical switching by single photons'',
	Nature Photonics, \textbf{6}, 607 (2012)
	
	
	\bibitem{Cancellieri}
	E. Cancellieri, A. Hayat, A. M. Steinberg, E. Giacobino,
	and A. Bramati, 
	``Ultrafast Stark-Induced Polaritonic Switches'',
	Phys. Rev. Lett., \textbf{112}, 053601 (2014).
	
	\bibitem{Heinze 2015}
	D. Heinze, D. Breddermann, A. Zrenner, and S. Schumacher, 
	``A quantum dot single-photon source with on-the-fly all-optical polarization control and timed emission'', Nature communications \textbf{6}, 8473 (2015).
	
	\bibitem{Nomura 2010}
	M. Nomura, N. Kumagai, S. Iwamoto, Y. Ota, and Y. Arakawa, 
	``Laser oscillation in a strongly coupled single-quantum-dot–nanocavity system'', Nature Physics, \textbf{6}, 279 (2010).
	
	
	\bibitem{hughes 2013}
	S. Hughes and H. J. Carmichael, 
	``Phonon-mediated population inversion in a semiconductor quantum-dot cavity system'',
	New J. Phys., \textbf{15}, 053039, (2013).
	
	\bibitem{Dara 2013}
	Dara P. S. McCutcheon and Ahsan Nazir,
	``Model of the Optical Emission of a Driven Semiconductor Quantum Dot: Phonon-Enhanced Coherent Scattering and Off-Resonant Sideband Narrowing'', Phys. Rev. Lett. \textbf{110}, 217401, (2013).
	
	\bibitem{Wei 2014}
	Yu-Jia Wei, Yu He, Yu-Ming He, Chao-Yang Lu, Jian-Wei Pan, Christian Schneider, Martin Kamp, Sven H\"ofling, Dara P. S.McCutcheon, and Ahsan Nazir, 
	``Temperature-Dependent Mollow Triplet Spectra from a Single Quantum Dot: Rabi Frequency Renormalization and Sideband Linewidth Insensitivity'', Phys. Rev. Lett. \textbf{113}, 097401, (2014).
	
	\bibitem{Muller 2015}
	Kai M\"uller, Kevin A. Fischer, Armand Rundquist, Constantin Dory, Konstantinos G. Lagoudakis, Tomas Sarmiento, Yousif A. Kelaita, Victoria Borish, and Jelena Vu\v{c}kovi\'c.,
	``Ultrafast Polariton-Phonon Dynamics of Strongly Coupled Quantum Dot-Nanocavity Systems'',
	Phys. Rev. X \textbf{5} 031006, (2015).
	
	\bibitem{Jake 2016}
	Jake Iles-Smith and Ahsan Nazir,
	``Quantum correlations of light and matter through environmental transitions'',
	Optica \textbf{3.2}, 207 (2016).
	
	\bibitem{Hargart 2016}
	F. Hargart, M. Müller, K. Roy-Choudhury, S. L. Portalupi, C.Schneider, S. Höfling, M. Kamp, S. Hughes, and P. Michler,
	``Cavity-enhanced simultaneous dressing of quantum dot exciton and biexciton states''
	Phys. Rev. B \textbf{93}, 115308, (2016).
	
\end{thebibliography}
\end{document}